\documentclass[aps,prl,twocolumn,superscriptaddress,longbibliography]{revtex4-2}
\usepackage{graphicx}
\usepackage{amssymb,amsmath}
\usepackage{bm}
\usepackage{dcolumn}
\usepackage{float}
\usepackage[OT1]{fontenc} 
\usepackage{url}
\usepackage{mathrsfs}
\usepackage{slashed,comment}
\usepackage{color}
\usepackage{verbatim}
\usepackage{enumitem}
\usepackage{soul,physics}
\usepackage[driverfallback=dvipdfm]{hyperref}
\hypersetup{pdfpagemode=FullScreen,colorlinks=true,breaklinks,urlcolor=blue,linkcolor=blue,citecolor=blue}

\usepackage{subfigure}
\usepackage{amssymb}
\usepackage{bm}
\usepackage{graphicx}
\usepackage{amsmath,amssymb}

\usepackage{pifont}
\def\X#1{%
        \raisebox{.9pt}{\textcircled{\raisebox{-.9pt}{#1}}}%
}

\usepackage{pdfpages} 
\usepackage{pgffor} 

\makeatletter
\AtBeginDocument{\let\LS@rot\@undefined}
\makeatother

\def\supplementfilename{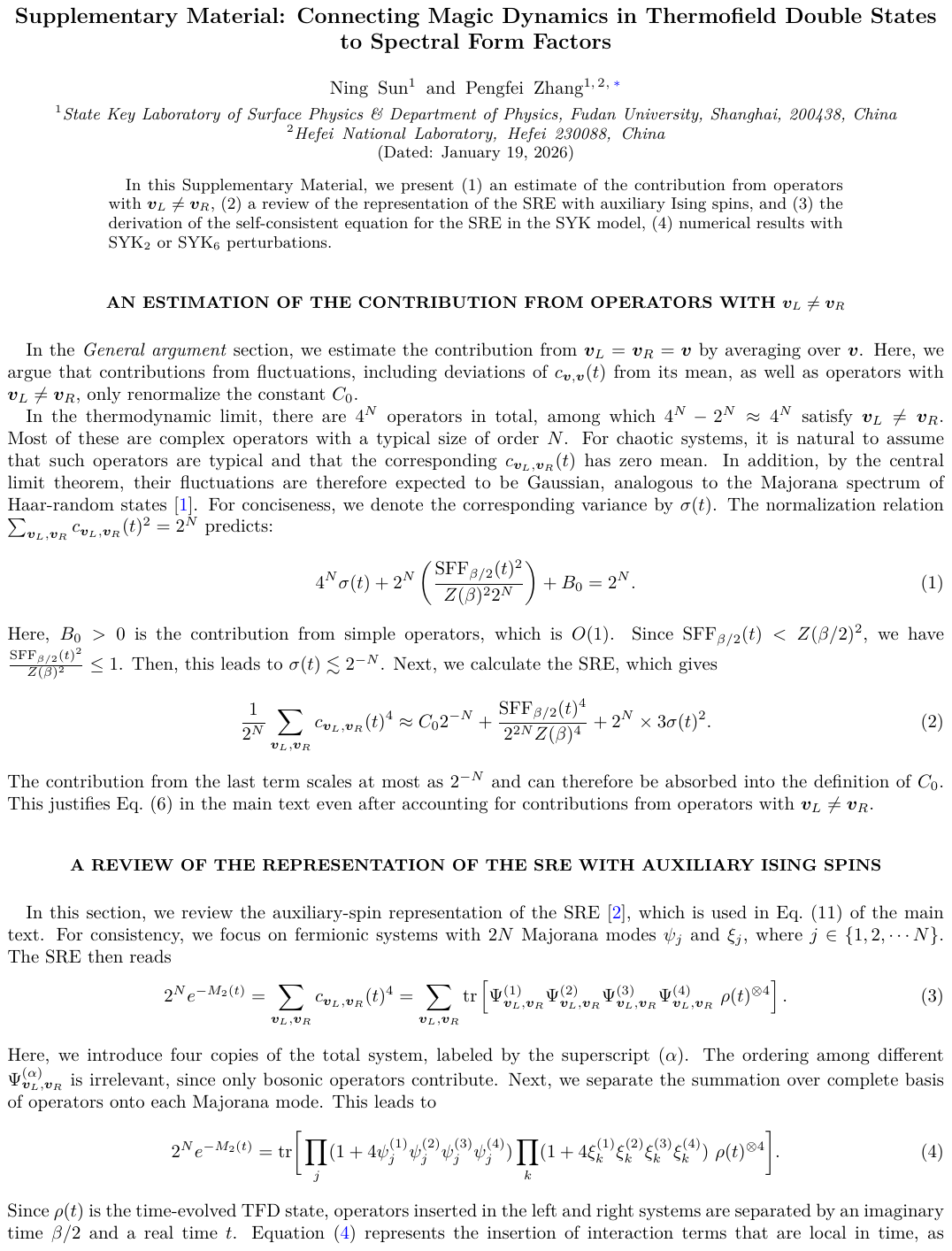}

\pdfximage{\supplementfilename}
\def\numbersupplementpages{\the\pdflastximagepages}

\newif\ifarXiv
\arXivtrue 

\begin{document}
 
  \title{Connecting Magic Dynamics in Thermofield Double States to Spectral Form Factors }

  \author{Ning Sun}
  \affiliation{State Key Laboratory of Surface Physics \& Department of Physics, Fudan University, Shanghai, 200438, China}

  \author{Pengfei Zhang}
  \thanks{PengfeiZhang.physics@gmail.com}
  \affiliation{State Key Laboratory of Surface Physics \& Department of Physics, Fudan University, Shanghai, 200438, China}
  \affiliation{Hefei National Laboratory, Hefei 230088, China}

  \date{\today}

  \begin{abstract}
  Under unitary evolution, chaotic quantum systems initialized in simple states rapidly develop high complexity, precluding any efficient classical description. Quantum chaos is traditionally characterized by spectral properties of the Hamiltonian, most notably through the spectral form factor, while the hardness of classical simulation within the stabilizer formalism, commonly referred to as quantum magic, can be quantified by the stabilizer R\'enyi entropy. In this Letter, we propose a relation between the dynamics of the stabilizer R\'enyi entropy for thermofield double states and the spectral form factor, based on general arguments for chaotic systems with all-to-all interactions. This relation implies that the saturation of the stabilizer R\'enyi entropy is governed by a first-order dynamical transition. We then demonstrate this relation explicitly in the Sachdev-Ye-Kitaev model, using an auxiliary-spin representation of the stabilizer R\'enyi entropy that exhibits an emergent $Z_2$ symmetry. We further find that, in the high-temperature regime of the SYK model, the transition occurs at a finite time, with the long-time phase marked by spontaneous $Z_2$ symmetry breaking. In contrast, at low temperatures, the transition is pushed to times exponentially long in the system size. Our results reveal an intriguing interplay between quantum chaos and quantum magic.
  \end{abstract}
    
  \maketitle

  \emph{ \color{blue}Introduction.---} While the Hilbert space dimension of quantum systems grows exponentially, certain physically relevant states still admit efficient classical representations. For example, states with low entanglement entropy can be described by tensor network states~\cite{White:1992zz,Orus:2018dya}, enabling the development of classical algorithms for computing ground-state properties of gapped quantum many-body systems. Another important class of quantum states consists of stabilizer states, which have wide applications in quantum error correction~\cite{nielsen2010quantum}. By the Gottesman-Knill theorem, such states can be efficiently represented within the stabilizer formalism~\cite{Gottesman:1997zz,Gottesman:1998hu,PhysRevA.57.127,PhysRevA.70.052328}. To quantify the distance between a given quantum state and the set of stabilizer states, the stabilizer R\'enyi entropy (SRE) has been introduced as a measure of non-stabilizerness, or quantum magic~\cite{Leone:2021rzd,Haug:2023hcs,Wang:2023uog,Huang:2025cbk,Xiao:2026ptn}. Alternative measures of quantum magic have also been proposed, including the robustness of magic, Wigner negativity, and mana~\cite{2017PhRvL.118i0501H,Liu:2020yso,Beverland:2019jej,PhysRevA.109.L040401,Turkeshi:2023ctq,msm2-vmg7,Ahmadi:2022bkg,Qian:2025oit}.

  More recently, considerable attention has been devoted to the study of quantum magic in quantum many-body systems, including Hamiltonian systems~\cite{White:2020zoz,Smith:2024ydp,Bera:2025pfp,Jasser:2025myz,njgn-fksh,Malvimat:2026oqf,Odavic:2024lqq,PhysRevA.110.022436,Tarabunga:2023xmv,Tarabunga:2023hau,Catalano:2024bdh,Viscardi:2025vya,Ding:2025nua,PRXQuantum.4.040317,Oliviero:2022euv,Collura:2024ida,Wang:2025csz,Li:2025rzd,Zhang:2025rky}, matrix product states~\cite{Lami:2023naw,PhysRevB.107.035148,PhysRevLett.133.010602,Tarabunga:2025wym}, and dynamics generated by quantum circuits~\cite{Turkeshi:2023lqu,Zhang:2024fyp,Turkeshi:2024pnj,Tirrito:2024kts,Haug:2024ptu,Szombathy:2024tow,Szombathy:2025euv,Hou:2025bau}. One fundamental question is how the SRE develops and saturates under chaotic quantum dynamics. Although the rapid growth of quantum magic is often associated with chaotic evolution, a precise and general connection between magic growth and the spectral statistics of the Hamiltonian~\cite{mehta2004random,Cotler:2016fpe,Saad:2018bqo}--diagnostics of quantum chaos--has not yet been established. Addressing this question provides important insight into the growth of complexity in chaotic systems, which underlies quantum thermalization~\cite{PhysRevA.43.2046,1999JPhA...32.1163S,2008Natur.452..854R,Kaufman:2016mif}. A major obstacle is the intrinsic hardness of classical simulation once the SRE becomes extensive, which prevents the identification of universal relations that emerge in the thermodynamic limit.

  In this Letter, we establish a direct relation between the dynamics of the SRE and the spectral form factor (SFF)~\cite{Cotler:2016fpe,Saad:2018bqo,Khramtsov:2020bvs} by focusing on thermofield double (TFD) states~\cite{Israel:1976ur,Maldacena:2001kr}, which have broad applications in condensed matter physics~\cite{Liu:2024mme,Weinstein:2024fug}, quantum information~\cite{Gao:2016bin,Maldacena:2017axo,Susskind:2017nto,Gao:2018yzk,Brown:2019hmk,Gao:2019nyj,Schuster:2021uvg,Jafferis:2022crx,Zhou:2024osg,Liu:2024nhs}, and gravity physics~\cite{Maldacena:2001kr}. We first present a general argument for this relation in chaotic systems with all-to-all interactions, and then demonstrate it explicitly in the Sachdev-Ye-Kitaev (SYK) model~\cite{Sachdev:1992fk,kitaev2014talk,maldacena2016remarks,RevModPhys.94.035004} using a path-integral approach with auxiliary spins that exhibit an emergent $Z_2$ symmetry~\cite{Zhang:2025rky}. This relation implies that the saturation of the SRE is accompanied by a first-order dynamical transition, producing a singularity in the SRE. In the SYK model, the short- and long-time dynamics are governed by distinct saddle points. The short-time saddle preserves the $Z_2$ symmetry, in which the SRE is related to the slope of the SFF, whereas the long-time saddle spontaneously breaks the $Z_2$ symmetry, yielding a nearly maximal SRE. At high temperatures, the transition occurs at a finite time; at low temperatures, the transition is controlled by soft modes and occurs only at exponentially long times. Our results uncover a fundamental link between quantum chaos and quantum magic.

  \emph{ \color{blue}Setup.---} We consider systems consisting of $N$ Majorana fermion modes $\psi_j$ that satisfy the canonical anticommutation relations $\{\psi_j,\psi_k\}=\delta_{jk}$. The Hamiltonian $H$ is assumed to be chaotic with all-to-all interactions, exhibiting random-matrix-like behavior in its level statistics~\cite{mehta2004random,Cotler:2016fpe}. A standard diagnostic of quantum chaos is the spectral form factor (SFF)~\cite{Cotler:2016fpe,Saad:2018bqo,Khramtsov:2020bvs}, defined as $\mathrm{SFF}_\beta(t)=\big|Z\left(\beta+it\right)\big|^2$. Here, $Z(\beta)=\mathrm{tr}\left(e^{-\beta H}\right)$ denotes the partition function of the thermal density matrix. For chaotic Hamiltonians, the SFF exhibits a characteristic slope-ramp-plateau structure as a function of time. At early times, the SFF shows a rapid decay, which is self-averaged~\cite{Saad:2018bqo}. At intermediate times, the SFF develops a linear ramp, which reflects level repulsion and spectral rigidity captured by random matrix theory. Finally, at late times beyond the Heisenberg time, the SFF saturates to a constant plateau at $Z(2\beta)$, signaling the discreteness of the energy spectrum. 

  We are interested in the quench dynamics of the system. In particular, we focus on a special class of pure initial states that admit a concrete path-integral representation, namely, TFD states~\cite{Israel:1976ur,Maldacena:2001kr}. These states are widely used to characterize symmetries of density matrices~\cite{Liu:2024mme,Weinstein:2024fug} and to perform many-body teleportation protocols~\cite{Gao:2016bin,Maldacena:2017axo,Susskind:2017nto,Gao:2018yzk,Brown:2019hmk,Gao:2019nyj,Schuster:2021uvg,Jafferis:2022crx,Zhou:2024osg,Liu:2024nhs}. To construct the TFD state, we introduce $N$ auxiliary fermion modes $\xi_j$ and define a maximally entangled state by the condition $(\psi_j+i\xi_j)\lvert\mathrm{EPR}\rangle=0$ for all $j$. Following the standard terminology, we refer to $\psi$ as the left system and $\xi$ as the right system. The TFD state is then defined as~\cite{Gu:2017njx}
  \begin{equation}
  |\text{TFD}\rangle=e^{-\frac{\beta}{2}H}\lvert\mathrm{EPR}\rangle/\sqrt{Z(\beta)},
  \end{equation}
  We initialize the total system with $2N$ Majorana fermions in the TFD state at $t=0$ and evolve it under the Hamiltonian $H$. At time $t$, the density matrix is $\rho(t)=e^{-iHt}\lvert\mathrm{TFD}\rangle\langle\mathrm{TFD}\rvert e^{iHt}$, for which we compute the SRE. In particular, the evolved state with $\beta=0$ can be viewed as the Choi representation of the unitary operator~\cite{Choi:1975nug}, which has been studied in~\cite{Hou:2025bau}. 

  The definition of the SRE in fermionic systems requires introducing an orthonormal operator basis of Majorana strings~\cite{Bera:2025pfp}:
  \begin{equation}
  \Psi_{\bm{v}_L,\bm{v}_R}=i^{\frac{v_t(v_t-1)}2}2^{\frac{v_t}{2}}\psi^{v_{L,1}}_1...\psi^{v_{L,N}}_{N}\xi^{v_{R,1}}_1...\xi^{v_{R,N}}_{N},
  \end{equation}
  where we have introduced two $N$-dimensional vectors $\bm{v}_L$ and $\bm{v}_R$, with components $v_{L,j}, v_{R,j}\in\{0,1\}$. The quantity $v_t=|\bm{v}_L|+|\bm{v}_R|\equiv \sum_j (v_{L,j}+v_{R,j})$ counts the number of nontrivial Majorana operators, and the coefficients ensure that $\Psi_{\bm{v}_L,\bm{v}_R}$ is Hermitian and satisfies $\Psi_{\bm{v}_L,\bm{v}_R}^2=1$. The density matrix can be expanded in this basis as $\rho(t)= \frac{1}{2^N}\sum_{{\bm{v}_L,\bm{v}_R}}c_{\bm{v}_L,\bm{v}_R}(t)\Psi_{\bm{v}_L,\bm{v}_R},$
  where $c_{\bm{v}_L,\bm{v}_R}(t)=\text{tr}_{LR}\left(\rho(t)\Psi_{\bm{v}_L,\bm{v}_R}\right)$ is known as the Majorana spectrum~\cite{Bera:2025pfp}. Here, we add a subscript to denote the trace over the total Hilbert space. The Majorana spectrum satisfies the normalization condition $\sum_{{\bm{v}_L,\bm{v}_R}}c_{\bm{v}_L,\bm{v}_R}(t)^2=2^N$. The SRE is then defined as 
  \begin{equation}\label{eq:defM2}
  M_2(t)\equiv -\ln \bigg(2^{-N}\sum_{{\bm{v}_L,\bm{v}_R}}c_{\bm{v}_L,\bm{v}_R}(t)^4\bigg).
  \end{equation}
  It measures the distance between a generic state and the set of stabilizer states. Any pure stabilizer state is a simultaneous eigenstate of $2^N$ commuting Majorana strings, known as the stabilizers. For these stabilizers, we have $|c_{\bm{v}_L,\bm{v}_R}(t)|=1$ while $|c_{\bm{v}_L,\bm{v}_R}(t)|=0$ for all remaining Majorana strings. An example is the maximally entangled state $\lvert\mathrm{EPR}\rangle$, which is stabilized by $2^N$ operators satisfying $\bm{v}_L=\bm{v}_R$. Consequently, we find $M_2=0$ for stabilizer states. For more general states, $M_2(t)$ satisfies the bound $0\leq M_2(t)\leq N\ln 2$~\cite{Leone:2021rzd}. In the following discussion, we focus primarily on the extensive part of the SRE, $\lim_{N\rightarrow \infty} M(t)/N$, neglecting subleading corrections arising from finite-size effects. 

  \begin{figure}[t]
    \centering
    \includegraphics[width=0.95\linewidth]{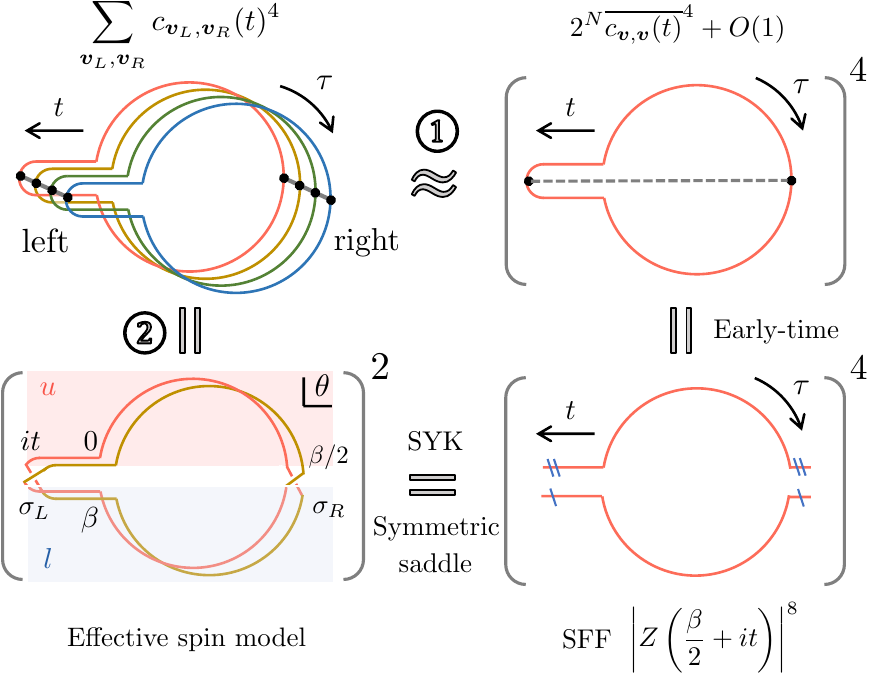}
    \caption{An illustration of two independent approaches to relate the SRE to the SFF at early times, as elaborated in the main text. In method~\X1, we focus on $c_{\bm{v},\bm{v}}(t)$ and approximate it by averaging over $\bm{v}$. The early-time regime before saturation is dominated by the contribution from the SFF. In method \X2, we focus on SYK-like models and assume that the $Z_2$ symmetry is preserved. In the long-time limit, saddles that break this symmetry can emerge.    }
    \label{fig:illustration}
  \end{figure}

 \emph{ \color{blue}General argument.---} We first present a general argument establishing the relation between the SRE and the SFF. Using properties of the EPR state, the Majorana spectrum can be expressed as a Wightman function 
 \begin{equation}\label{eq:Wightman}
 c_{\bm{v}_L,\bm{v}_R}(t)=\frac{i^{n_{\bm{v}_L,\bm{v}_R}}}{Z(\beta)}\text{tr}\Big(e^{-(\frac{\beta}{2}-it) H}\Psi_{\bm{v}_L,\bm{0}}e^{-(\frac{\beta}{2}+it) H}\Psi_{\bm{v}_R,\bm{0}}\Big).
 \end{equation}
 Here, for conciseness, we introduce the integer $n_{\bm{v}_L,\bm{v}_R}=(|\bm{v}_L|+1)|\bm{v}_R|$. For chaotic systems, the Wightman function is expected to decay with time. On the other hand, at $t=\beta=0$ we have $c_{\bm{v}_L,\bm{v}_R}(t)=i^{n_{\bm{v}_L,\bm{v}_R}}$ for $\bm{v}_L=\bm{v}_R$, and $c_{\bm{v}_L,\bm{v}_R}(t)=0$ otherwise. It is therefore natural to focus on the contribution from $c_{\bm{v},\bm{v}}(t)$ in Eq.~\eqref{eq:defM2}. Next, we classify operators into three categories~\footnote{This simple classification neglects the locality structure of the model and is therefore expected to be valid for systems with all-to-all interactions.}: (1) operators with small size $|\bm{v}|$, (2) operators with near-maximal size $|\bm{v}|\approx N$, and (3) complex operators with $|\bm{v}|,\ N-|\bm{v}|\gg 1$. Class (1) contains simple operators, for which the Wightman function is finite; however, the number of such operators is itself only $O(1)$. Operators in class (2) are analogous, as they are obtained by multiplying class (1) operators by the fermion parity operator. By contrast, the overwhelming majority of operators belong to class (3). To estimate the contribution from this large set, we average over all choices of $\bm{v}$, leading to
\begin{equation}\label{eq:averagedc}
 \overline{(-i)^{n_{\bm{v},\bm{v}}}c_{\bm{v},\bm{v}}(t)}=\frac{1}{Z(\beta)2^{N/2}}\left|Z\left(\tfrac{\beta}{2}+it\right)\right|^2.
 \end{equation}
 Here, we isolate a phase factor. The SFF already appears. Neglecting the fluctuations of $c_{\bm{v}_L,\bm{v}_R}(t)$ around its expectation value and adding back contributions from simple operators, we find 
 \begin{equation}\label{eq:sum}
 \frac{1}{2^{N}}\sum_{{\bm{v}_L,\bm{v}_R}}c_{\bm{v}_L,\bm{v}_R}(t)^4\approx C_0 2^{-N}+\frac{\text{SFF}_{{\beta}/{2}}(t)^4}{2^{2N}Z(\beta)^4}.
 \end{equation}
 Here, $C_0\sim O(1)$ is a constant. We expect that contributions from operators with $\bm{v}_L\neq\bm{v}_R$ only renormalize $C_0$, as estimated in the Supplementary Material~\cite{SM}.
 
 Eq.~\eqref{eq:sum} shows the expected behavior of magic growth. At $\beta=t=0$, the second term becomes unity while the first term is negligible, leading to $M_2=0$. With this motivation, one may try dropping the first term:
 \begin{equation}\label{eq:early}
 M_2(t)\approx 2N\ln 2-4\ln \left[\text{SFF}_{{\beta}/{2}}(t)/Z(\beta)\right]\equiv M_2^{(p)}(t).
 \end{equation}
 This provides a direct relation between the SRE and the SFF. Using the general features of the SFF, Eq.~\eqref{eq:early} predicts a nonmonotonic behavior of the SRE. In the slope regime of the SFF, the SRE increases and is self-averaging. This increase terminates when the dynamics enter the ramp regime, where the SRE decays logarithmically. Finally, the SRE saturates in the plateau regime. However, since $\mathrm{SFF}_{\beta/2}(\infty)=Z(\beta)$ in the plateau regime, Eq.~\eqref{eq:early} predicts $M_2(t)=2N\ln 2$ in the long-time limit. This indicates that Eq.~\eqref{eq:early} already violates the bound $M_2(t)\leq N\ln 2$ in the slope regime at sufficiently long times. This violation is a close analog of the information paradox for the entanglement entropy~\cite{PhysRevD.14.2460}, particularly for the dynamics of the thermofield double state~\cite{Almheiri:2019yqk,Chen:2020wiq}. Adding back the $2^{-N}$ term resolves this paradox and, in the thermodynamic limit, leads to
 \begin{equation}\label{eq:full}
  M_2(t)\approx \text{Min}\{N\ln 2, M_2^{(p)}(t)\}.
 \end{equation}
 For high-temperature regime, this results in a first-order dynamical transition at $t_*$ (in the slope regime) determined by $N\ln 2=M_2^{(p)}(t_*)$, resulting a singularity in the SRE. In the low-temperature limit, it is possible that $M_2^{(p)}(t) > N\ln 2$ for arbitrary times $t$, due to the extensive magic generated by imaginary-time evolution.

 \emph{ \color{blue}The Sachdev-Ye-Kitaev model.---} To further demonstrate the validity of Eq.~\eqref{eq:full} in microscopic Hamiltonians without introducing Haar integrals, we study the dynamics of the SRE in the SYK model~\cite{Sachdev:1992fk,kitaev2014talk,maldacena2016remarks,RevModPhys.94.035004}. The model describes $N$ randomly interacting Majorana fermion modes with Hamiltonian
\begin{equation}\label{eqn:H}
  H=\sum_{i_1<i_2<\cdots<i_q}i^{\frac{q(q-1)}2}J_{i_1i_2\cdots i_q}\psi_{i_1}\psi_{i_2}\cdots\psi_{i_q}, 
  \end{equation}
  where the random couplings $J_{i_1i_2\cdots i_q}$ are independent Gaussian variables with 
  \begin{equation}
  \overline{J_{i_1i_2\cdots i_q}}=0,\ \ \ \ \ \ \overline{J_{i_1i_2\cdots i_q}^2}=(q-1)!J^2/N^{q-1}. 
  \end{equation}
  The SYK model is chaotic for $q \ge 4$, exhibiting random-matrix behavior~\cite{Cotler:2016fpe} and an emergent holography~\cite{Maldacena:2016upp,maldacena2016remarks,kitaev2014talk}. It also serves as a concrete solvable model for non-Fermi liquids~\cite{RevModPhys.94.035004}. Numerical studies of the SRE dynamics in the SYK model have been performed using exact diagonalization for finite $N$ in Refs.~\cite{Bera:2025pfp,Jasser:2025myz,njgn-fksh,Malvimat:2026oqf}. 

  We study the SRE dynamics of the SYK model in TFD states at $N\rightarrow \infty$ by employing the path-integral representation of $Z_{\text{SRE}}=Z(\beta)^4\sum_{{\bm{v}_L,\bm{v}_R}}c_{\bm{v}_L,\bm{v}_R}(t)^4$, as illustrated in Fig.~\ref{fig:illustration}. Four copies of the Majorana fermion fields $\psi_j^{(\alpha)}$, with $\alpha \in \{1,2,3,4\}$, arise from the four replicas of the density matrix (different colors in Fig.~\ref{fig:illustration}). The original definition of the SRE requires extensive operator insertions. Following the construction in Ref.~\cite{Zhang:2025rky}, this can be equivalently formulated by introducing auxiliary Ising spins~\cite{SM}, which leads to 
  \begin{equation}
  \begin{aligned}
  Z_{\text{SRE}}=&\sum_{\sigma_{L/R,j}}\Big< \prod_j\mathcal{S}_{j,it}^{(12,34)}[\sigma_{L,j}]\mathcal{S}_{j,\beta/2}^{(12,34)}[\sigma_{R,j}] \Big>_0,
  \end{aligned}
  \end{equation}
  where $\sigma_{L/R,j}=\pm 1$ are $2N$ Ising variables. $\mathcal{S}_{j,\theta}^{(12,34)}[\sigma]$ is the fermionic SWAP (fSWAP) operation (with complex time $\theta=\tau+it$) that imposes
  \begin{equation}
  \psi_j^{(1)}(\theta^+)=-\sigma\psi_j^{(2)}(\theta^-),\ \ \ \psi_j^{(2)}(\theta^+)=\sigma\psi_j^{(1)}(\theta^-),
  \end{equation}
  and similarly for $(\psi_j^{(3)},\psi_j^{(4)})$. We highlight that the effective spin model now exhibits a global $Z_2$ symmetry $(\sigma_{L/R,j},\psi^{(2)}_j,\psi^{(4)}_j)\rightarrow -(\sigma_{L/R,j},\psi^{(2)}_j,\psi^{(4)}_j)$, which plays an central role in later discussions.

   Due to the presence of two fSWAP operators for each pair of replicas, the upper branch (denoted as $u$) of one replica becomes connected to the lower branch (denoted as $l$) of the other replica, with an additional phase $\sigma_{L/R,j}$ acquired when passing through the connection point, as illustrated in Fig.~\ref{fig:illustration}. Therefore, it is convenient to define
   \begin{equation}
   \psi_j^{[1]}(\theta)\equiv \begin{cases} \psi_j^{(1)}(\theta) & \theta\in u\ \left(\text{Re~}\theta<\beta/2\right),\\
     \psi_j^{(2)}(\theta)  & \theta\in l\ \left(\text{Re~}\theta>\beta/2\right),
    \end{cases}
   \end{equation}
   and similarly for other replicas. Then, $\psi_j^{[\alpha]}$ with different $\alpha$ becomes disconnected. For the SYK model, this implies that the two-point function $G^{[\alpha\gamma]}(\theta_1,\theta_2)=\langle \psi^{[\alpha]}_j(\theta_1) \psi^{[\gamma]}_j(\theta_2)\rangle=\delta^{\alpha\gamma}G(\theta_1,\theta_2)$ is diagonal in $\alpha\gamma$~\cite{Kitaev:2017awl}. As derived in the Supplementary Material~\cite{SM}, the Green's function satisfies the saddle-point equations:
   \begin{equation}\label{eqn:saddle}
   \begin{aligned}
   \Sigma(\theta_1,\theta_2)&=J^2f(\theta_1)f(\theta_2)G(\theta_1,\theta_2)^{q-1},\\
   G(\theta_1,\theta_2)&=\frac{\sum_\sigma g_\sigma(\theta_1,\theta_2)[\text{det} g_\sigma]^{-2}}{\sum_\sigma [\text{det} g_\sigma]^{-2}}.
   \end{aligned}
   \end{equation}
   Here, we have introduced $f(\theta)=1,, i,,-i$ when $\theta$ lies on the branch with imaginary, forward, or backward evolution, respectively~\cite{Zhang:2020kia,SM}. The function $g_\sigma=(\partial_\tau-\Sigma)^{-1}_\sigma$ denotes the Green’s function evaluated with $\sigma_L=\sigma_R=\sigma$ for a representative site, where the $u$ and $l$ branches are joined with a phase $\sigma$ at both ends. The SRE can then be computed using the saddle-point solutions~\cite{SM}.

   We are now ready to demonstrate the relation between the SRE and the SFF in the SYK model. First, we assume that the $Z_2$ symmetry is unbroken. As a result, both $G(\theta_1,\theta_2)$ and $\Sigma(\theta_1,\theta_2)$ are nonzero only if $\theta_1$ and $\theta_2$ both lie on either the $u$ or the $l$ branch. The self-consistent equation can be equivalently written as 
   \begin{equation}
   \Sigma(\theta_1,\theta_2)=J^2P(\theta_1,\theta_2)f(\theta_1)f(\theta_2)g_\sigma(\theta_1,\theta_2)^{q-1}.
   \end{equation}
   Here, $P(\theta_1,\theta_2)=1$ if $\theta_1$ and $\theta_2$ both lie on the $u$ or the $l$ branch, and vanishes otherwise. It also guarantees that the saddle-point solution of $\Sigma(\theta_1,\theta_2)$ is independent of $\sigma$. On the other hand, the same saddle-point equation shows up when evaluating 
   \begin{equation}
   \tilde{Z}=\langle \text{EPR}_{\bm{\sigma}}|e^{-(\frac{\beta}{2}-it) H}\otimes e^{-(\frac{\beta}{2}+it) \tilde{H}}|\text{EPR}_{\bm{\sigma}}\rangle.
   \end{equation}
   This inner product is again defined for a system containing $2N$ Majorana fermions, $\psi_j$ and $\xi_j$. The vector $\bm{\sigma}$ is an $N$-dimensional vector that specifies the EPR state via $(\psi_j + i \sigma_j \xi_j)\lvert \mathrm{EPR}_{\bm{\sigma}} \rangle = 0$. The Hamiltonian $H$ is given by Eq.~\eqref{eqn:H}, while $\tilde{H}$ is obtained from $H$ by (1) replacing $\psi_j$ with $\xi_j$ and (2) replacing $J_{i_1 i_2 \cdots i_q}$ with independently sampled couplings $\tilde{J}_{i_1 i_2 \cdots i_q}$. In particular, the presence of independent couplings guarantees the appearance of $P(\theta_1,\theta_2)$ in the saddle-point equations. Since the partition function of the SRE contains four replicas, we expect $Z_{\text{SRE}} \propto \tilde{Z}^4$, up to factors that are independent of $\beta J$ and $tJ$. Finally, because $\tilde{Z}$ is independent of $\bm{\sigma}$, we can perform a summation over all $\bm{\sigma}$. Noting that the states $\lvert \mathrm{EPR}_{\bm{\sigma}} \rangle$ with different $\bm{\sigma}$ form a complete basis, we find
   \begin{equation}\label{eq:SYKrelation}
   \tilde{Z}\propto\text{tr}\left[e^{-(\frac{\beta}{2}-it) H}\right]\times \text{tr}\left[e^{-(\frac{\beta}{2}+it) \tilde{H}}\right]=\text{SFF}_{{\beta}/{2}}(t)\Big|_{\text{slope}}.
   \end{equation}
   This relation between $\tilde{Z}$ and the SFF holds only when the SFF is dominated by the disconnected solution between the forward- and backward-evolution branches. This precisely corresponds to the self-averaged slope regime~\cite{Saad:2018bqo}. Using Eq.~\eqref{eq:SYKrelation}, we conclude that the relation proposed in Eq.~\eqref{eq:early} becomes \textit{exact} for the symmetric saddle of the SRE and the slope regime of the SFF. 

   The discussion in the previous section indicates that the symmetric saddle is eventually replaced by a saddle exhibiting spontaneous symmetry breaking (SSB), which prevents the growth of the SRE beyond its upper bound. Here, we focus exclusively on the regime deep in the symmetry-breaking phase, where only $\sigma=1$ contributes in Eq.~\eqref{eqn:saddle} due to $\text{det} g_1\ll \text{det} g_{-1}$. As a result, we can approximate $G(\theta_1,\theta_2)\approx g_1(\theta_1,\theta_2)$, and the saddle-point equation reduces to that of the conventional Keldysh contour, $Z(\beta)\approx\text{tr}[e^{iHt}e^{-iHt}e^{-\beta H}]$~\cite{kamenev2023field}. Accounting for the four replicas, we then obtain $Z_{\text{SRE}} = {Z}(\beta)^4$, which yields $M_2(t)\approx N\ln 2$. Combined with the symmetric saddle, this yields the behavior in Eq.~\eqref{eq:full}. Finally, we emphasize that although the above discussion is illustrated using the original SYK model, the same conclusions apply to a large class of generalizations such as randomly coupled SYK models~\cite{Banerjee:2016ncu,Chen:2017dav,PhysRevLett.119.216601,PhysRevLett.119.206602} or higher-dimensional SYK models~\cite{Gu:2016oyy,PhysRevB.95.155131}.

  \begin{figure}[t]
    \centering
    \includegraphics[width=0.99\linewidth]{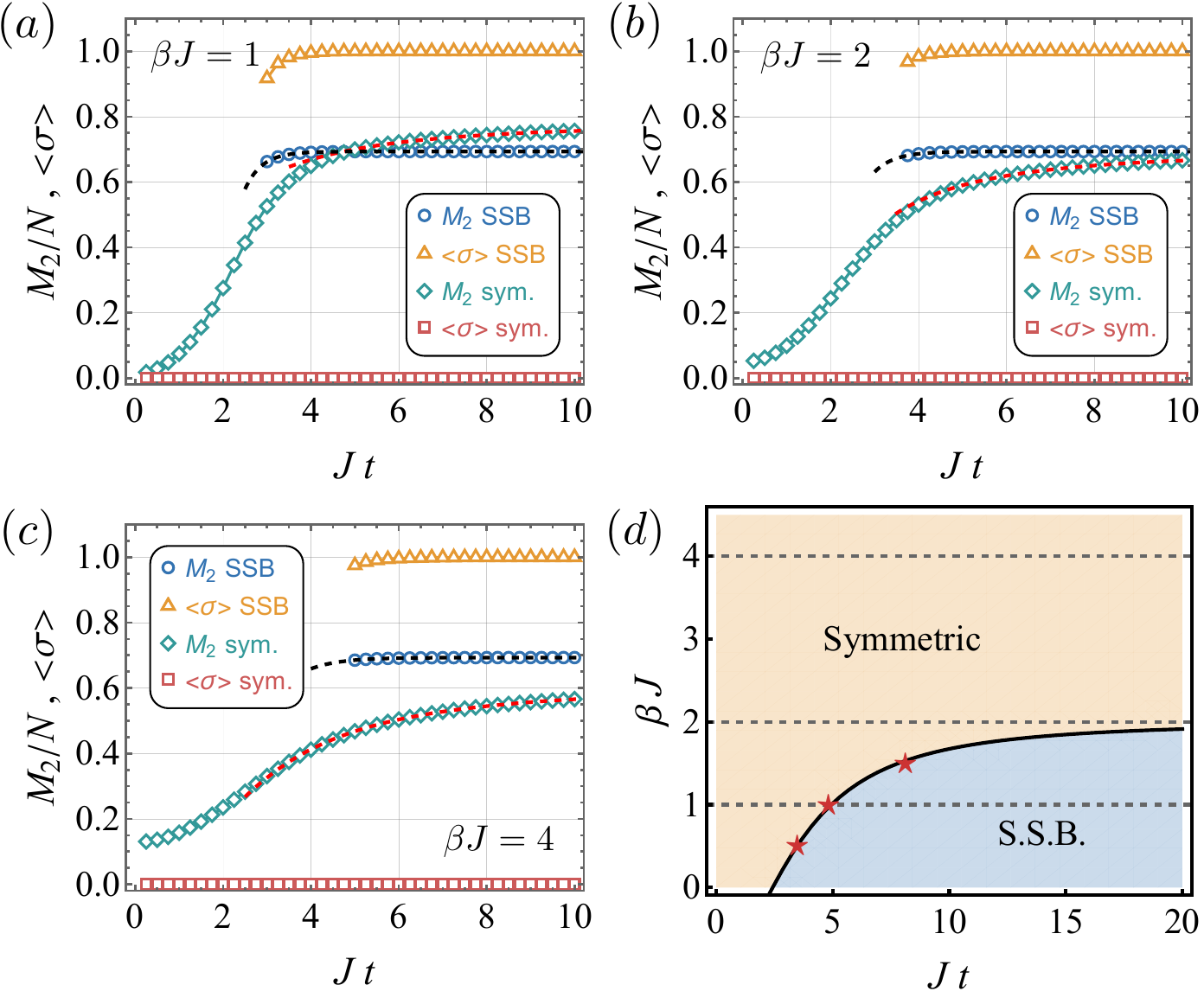}
    \caption{Numerical results for the SRE in the SYK model with $q=4$. In panels (a-c), we present the time evolution of the SRE and the order parameter $\langle \sigma\rangle$ on different saddle points at different temperatures. The dashed lines represent fits, as described in the main text. In panel (d), we show the phase diagram of the SRE. The phase boundary is obtained by fitting data points denoted by red stars (with an additional data point at $(48.3,2)$). }
    \label{fig:num}
  \end{figure}

 \emph{ \color{blue}Numerics.---} We now present results for the SRE obtained by numerically solving the saddle-point equations for $q=4$. Similar methods have previously been applied to the study of the R\'enyi entropy~\cite{Chen:2020wiq,Zhang:2020kia}. As shown in FIG.~\ref{fig:num}(a-c), there exist multiple saddle-point solutions, which are plotted together with the order parameter $\langle \sigma\rangle$ for the $Z_2$ symmetry, defined as
 \begin{equation}
 \langle \sigma\rangle={\sum_\sigma\sigma~[\text{det} g_\sigma]^{-2}}\Big/{\sum_\sigma [\text{det} g_\sigma]^{-2}}.
 \end{equation} 
 The saddle-point solution with $\langle \sigma\rangle = 0$ is symmetric, while the solution with $\langle \sigma\rangle \neq 0$ exhibits SSB. We have also numerically verified that the symmetric solution satisfies Eq.~\eqref{eq:early} to machine precision by independently computing the SFF in the slope regime.

 At high temperatures, $\beta J = 1$, the early-time SRE is dominated by the symmetric saddle, while the dynamical transition to the saddle with SSB occurs at $t_* J \approx 4.8$. The symmetry-breaking saddle is nearly fully polarized, leading to an almost maximal SRE. To further quantify the late-time behavior, we fit the numerical results on the SSB saddle using $M_2/N = \ln 2 - a_0 e^{-b_0 J t}$. The resulting fit, shown by the black dashed lines, matches the numerical data with high accuracy. When we the temperature decreases, there are important modifications. First, the initial value $M_2(0)$ increases due to the imaginary-time evolution. The Eq.~\eqref{eq:early} predicts $M_2(0)=2N\ln 2-4S^{(2)}(\beta/2)$, where $S^{(2)}(\beta)$ is the second R\'enyi entropy of the thermal density matrix. For the SYK model, $S^{(2)}(\infty)=0.2324$~\cite{maldacena2016remarks}. As a result, we always have $M_2(0)<N\ln 2$. Second, the growth of the symmetric solution becomes significantly slower and does not exhibit a transition to the SSB saddle at finite $t$ for $\beta J \gtrsim 2$. We expect that this behavior stems from the soft reparametrization mode in the SYK model, which similarly slows entanglement dynamics~\cite{Gu:2017njx,Sohal:2021djw}. These reparametrization modes are governed by the Schwarzian action~\cite{maldacena2016remarks}, which admits an exact partition function~\cite{Stanford:2017thb} $Z_\text{sch}(\beta)\propto \beta^{-3/2}\exp(\frac{2\pi^2 C}{ \beta J})$ with $C\propto N$. In the slope regime, the SFF can be obtained by the analytical continuation of $Z_\text{sch}$, which leads to 
 \begin{equation}\label{eq:reparametrization}
  M_2^{(p)}(t)=M_2(0)+\frac{32\pi^2 C}{ \beta J}\frac{4t^2}{{\beta^2}+4t^2}+6\ln\left(\frac{{\beta^2}+4t^2}{\beta^2}\right).
 \end{equation}
 The saddle-point analysis reproduces only the first two terms, which are extensive in $N$. Consequently, we obtain $M_2(\infty)-M_2(0)=\frac{32\pi^2 C}{\beta J}$,
which decreases as $\beta J$ increases, consistent with the numerical observations. Further numerical evidence for the relevance of soft modes is provided in the Supplementary Material~\cite{SM}. Motivated by this analysis, we further fit the numerical results for the symmetric saddle using $M_2/N = a_0 + \frac{b_0}{c_0 + J^2t^2}$. The resulting fits, shown by the red dashed lines, match the numerical data at long times with good accuracy even for $\beta J=1$. The analysis also indicates that at exponential long times before the dip time, $t \sim e^{N}\lesssim t_{\text{dip}}$, the last term in Eq.~\eqref{eq:reparametrization} yields $M_2^{(p)}(t) > N\ln 2$, at which point a transition to the symmetry-breaking saddle occurs. Finally, we present a phase diagram for the SRE in FIG.~\ref{fig:num}(d), where the phase boundary is determined by fitting the numerical data with $\beta J=a_0+\frac{b_0}{c_0+J^2t_*^2}$.

\emph{ \color{blue}Discussions.---} In this Letter, we establish a connection between the SRE dynamics of TFD states and the SFF through both a general argument and a concrete study of the SYK model. Our results reveal a dynamical transition between an early-time regime, captured by the slope regime of the SFF, and a late-time regime with a nearly maximal SRE. In the SYK model, these distinct regimes are described by different saddle points in the path-integral representation, where an emergent $\mathbb{Z}_2$ symmetry plays a pivotal role. At low temperatures, soft reparametrization modes delay this transition to exponentially long times. Our findings provide a universal link between spectral diagnostics of quantum chaos and the emergence of classical simulation hardness, shedding new light on complexity growth in thermalizing quantum systems.

We conclude with several remarks on future directions. While our analysis focuses on models without locality, it would be particularly interesting to test whether the same relation persists in general systems with finite-range interactions. Investigating higher-order R\'enyi entropies or alternative measures of quantum magic (e.g., mana) and their connections to the SFF may uncover richer structures and offer a more unified perspective on magic dynamics. Finally, it is natural to generalize the unitary dynamics considered here to monitored dynamics, where measurement-induced phase transitions can occur~\cite{PhysRevX.9.031009,PhysRevB.98.205136,PhysRevB.99.224307}. Such extensions may lead to analogs of our results in systems with non-unitary evolutions. We leave these directions for future work.

\vspace{5pt}
\textit{Acknowledgement.}
We thank Hanteng Wang, Zhi-Cheng Yang, Tian-Gang Zhou, and Yi-Neng Zhou for helpful discussions. This project is supported by the Shanghai Rising-Star Program under grant number 24QA2700300, the NSFC under grant 12374477, and the Quantum Science and Technology-National Science and Technology Major Project 2024ZD0300101.

\bibliography{main.bbl}

\ifarXiv


\section*{Supplementary material}

# Supplementary Material: Connecting Magic Dynamics in Thermofield Double States to Spectral Form Factors

Ning Sun[1] and Pengfei Zhang[1, 2, *]

[1] *State Key Laboratory of Surface Physics & Department of Physics, Fudan University, Shanghai, 200438, China*
[2] *Hefei National Laboratory, Hefei 230088, China*

(Dated: January 19, 2026)

In this Supplementary Material, we present (1) an estimate of the contribution from operators with $\bm{v}_L \neq \bm{v}_R$, (2) a review of the representation of the SRE with auxiliary Ising spins, and (3) the derivation of the self-consistent equation for the SRE in the SYK model, (4) numerical results with $\text{SYK}_2$ or $\text{SYK}_6$ perturbations.

## AN ESTIMATION OF THE CONTRIBUTION FROM OPERATORS WITH $\bm{v}_L \neq \bm{v}_R$

In the *General argument* section, we estimate the contribution from $\bm{v}_L = \bm{v}_R = \bm{v}$ by averaging over $\bm{v}$. Here, we argue that contributions from fluctuations, including deviations of $c_{\bm{v},\bm{v}}(t)$ from its mean, as well as operators with $\bm{v}_L \neq \bm{v}_R$, only renormalize the constant $C_0$.

In the thermodynamic limit, there are $4^N$ operators in total, among which $4^N - 2^N \approx 4^N$ satisfy $\bm{v}_L \neq \bm{v}_R$. Most of these are complex operators with a typical size of order $N$. For chaotic systems, it is natural to assume that such operators are typical and that the corresponding $c_{\bm{v}_L,\bm{v}_R}(t)$ has zero mean. In addition, by the central limit theorem, their fluctuations are therefore expected to be Gaussian, analogous to the Majorana spectrum of Haar-random states [1]. For conciseness, we denote the corresponding variance by $\sigma(t)$. The normalization relation $\sum_{\bm{v}_L,\bm{v}_R} c_{\bm{v}_L,\bm{v}_R}(t)^2 = 2^N$ predicts:

$$4^N \sigma(t) + 2^N \left(\frac{\text{SFF}_{\beta/2}(t)^2}{Z(\beta)^2 2^N}\right) + B_0 = 2^N. \tag{1}$$

Here, $B_0 > 0$ is the contribution from simple operators, which is $O(1)$. Since $\text{SFF}_{\beta/2}(t) < Z(\beta/2)^2$, we have $\frac{\text{SFF}_{\beta/2}(t)^2}{Z(\beta)^2} \leq 1$. Then, this leads to $\sigma(t) \lesssim 2^{-N}$. Next, we calculate the SRE, which gives

$$\frac{1}{2^N} \sum_{\bm{v}_L,\bm{v}_R} c_{\bm{v}_L,\bm{v}_R}(t)^4 \approx C_0 2^{-N} + \frac{\text{SFF}_{\beta/2}(t)^4}{2^{2N} Z(\beta)^4} + 2^N \times 3\sigma(t)^2. \tag{2}$$

The contribution from the last term scales at most as $2^{-N}$ and can therefore be absorbed into the definition of $C_0$. This justifies Eq. (6) in the main text even after accounting for contributions from operators with $\bm{v}_L \neq \bm{v}_R$.

## A REVIEW OF THE REPRESENTATION OF THE SRE WITH AUXILIARY ISING SPINS

In this section, we review the auxiliary-spin representation of the SRE [2], which is used in Eq. (11) of the main text. For consistency, we focus on fermionic systems with $2N$ Majorana modes $\psi_j$ and $\xi_j$, where $j \in \{1, 2, \cdots N\}$. The SRE then reads

$$2^N e^{-M_2(t)} = \sum_{\bm{v}_L,\bm{v}_R} c_{\bm{v}_L,\bm{v}_R}(t)^4 = \sum_{\bm{v}_L,\bm{v}_R} \text{tr}\left[\Psi^{(1)}_{\bm{v}_L,\bm{v}_R} \Psi^{(2)}_{\bm{v}_L,\bm{v}_R} \Psi^{(3)}_{\bm{v}_L,\bm{v}_R} \Psi^{(4)}_{\bm{v}_L,\bm{v}_R} \rho(t)^{\otimes 4}\right]. \tag{3}$$

Here, we introduce four copies of the total system, labeled by the superscript $(\alpha)$. The ordering among different $\Psi^{(\alpha)}_{\bm{v}_L,\bm{v}_R}$ is irrelevant, since only bosonic operators contribute. Next, we separate the summation over complete basis of operators onto each Majorana mode. This leads to

$$2^N e^{-M_2(t)} = \text{tr}\left[\prod_j (1 + 4\psi_j^{(1)} \psi_j^{(2)} \psi_j^{(3)} \psi_j^{(4)}) \prod_k (1 + 4\xi_k^{(1)} \xi_k^{(2)} \xi_k^{(3)} \xi_k^{(4)}) \rho(t)^{\otimes 4}\right]. \tag{4}$$

Since $\rho(t)$ is the time-evolved TFD state, operators inserted in the left and right systems are separated by an imaginary time $\beta/2$ and a real time $t$. Equation (4) represents the insertion of interaction terms that are local in time, as

illustrated in FIG. 1 of the main text. To proceed, we can perform a discrete version of the Hubbard-Stratonovich transformation, which takes the form of

$$(1 + 4\psi_j^{(1)}\psi_j^{(2)}\psi_j^{(3)}\psi_j^{(4)}) = \frac{1}{2}\sum_{\sigma_{L,j}=\pm 1}(1 + 2\sigma_{L,j}\psi_j^{(1)}\psi_j^{(2)})(1 + 2\sigma_{L,j}\psi_j^{(3)}\psi_j^{(4)}). \tag{5}$$

We further identify that $\mathcal{S}_{j,L}^{(12)}(\sigma_{L,j}) = \frac{1}{\sqrt{2}}(1 + 2\sigma_{L,j}\psi_j^{(1)}\psi_j^{(2)})$ is the fermion SWAP operator, since it is unitary $\mathcal{S}_{j,L}^{(12)}(\sigma_{L,j})\mathcal{S}_{j,L}^{(12)}(\sigma_{L,j})^\dagger = 1$, and satisfies

$$\mathcal{S}_{j,L}^{(12)}(\sigma_{L,j})\big(\psi_j^{(1)}, \psi_j^{(2)}\big)\mathcal{S}_{j,L}^{(12)}(\sigma_{L,j})^\dagger = \sigma_{L,j}\big(-\psi_j^{(2)}, \psi_j^{(1)}\big). \tag{6}$$

Putting all ingredients together, we find

$$2^N e^{-M_2(t)} = \sum_{\sigma_{L/R,j}} \text{tr}\bigg[\prod_j \mathcal{S}_{j,L}^{(12)}(\sigma_{L,j})\mathcal{S}_{j,L}^{(34)}(\sigma_{L,j})\mathcal{S}_{j,R}^{(12)}(\sigma_{R,j})\mathcal{S}_{j,R}^{(34)}(\sigma_{R,j})\ \rho(t)^{\otimes 4}\bigg]. \tag{7}$$

Finally, by expressing the expectation value using the standard path-integral formalism (see Ref. [3] for TFD states), we arrive at Eq. (11) in the main text, where the $L/R$ labels are replaced by the corresponding complex times $\theta = it$ or $\theta = \beta/2$:

$$\mathcal{S}_{j,it}^{(12,34)}[\sigma_{L,j}] = \mathcal{S}_{j,L}^{(12)}(\sigma_{L,j})\mathcal{S}_{j,L}^{(34)}(\sigma_{L,j}), \qquad \mathcal{S}_{j,\beta/2}^{(12,34)}[\sigma_{R,j}] = \mathcal{S}_{j,R}^{(12)}(\sigma_{R,j})\mathcal{S}_{j,R}^{(34)}(\sigma_{R,j}). \tag{8}$$

**THE DERIVATION OF THE SELF-CONSISTENT EQUATION FOR THE SRE IN THE SYK MODEL**

In this section, we derive the Schwinger-Dyson equations governing the dynamics of the SRE in the SYK model. We begin by writing Eq. (11) in a fully explicit form within the path-integral formalism, which reads

$$Z_{\text{SRE}} = \sum_{\{\sigma_{L/R,j}\}} \int dJ_{i_1 i_2 \cdots i_q} P(J_{i_1 i_2 \cdots i_q}) \int_{\text{B.C.}} D\psi_j^{(\alpha)}\ \exp\Big(-S[\psi_j^{(\alpha)}]\Big),$$

$$S = \int_0^{\beta+2t} ds \left[\sum_{j,\alpha}\frac{1}{2}\psi_j^{(\alpha)}\partial_s\psi_j^{(\alpha)} + \sum_\alpha f(s)H[\psi_j^{(\alpha)}]\right]. \tag{9}$$

Here, the $\xi_j$ operators are converted to $\psi_j$ operators using the definition of the EPR state and therefore do not appear explicitly. For convenience, we introduce a real parameter $s \in [0, \beta+2t)$ to label the entire Keldysh contour. The segment $s \in [0,t)$ corresponds to backward real-time evolution, $s \in [t, t+\beta)$ to imaginary-time evolution, and $s \in [t+\beta, \beta+2t)$ to forward real-time evolution. B.C. denotes the boundary condition (see Eq. (13) in the main text) induced by the twist operators, which depends on $\sigma_{L/R,j}$. Using our parametrization with $s$, it becomes

$$\psi_j^{(1)}(2t+\beta-0^+) = \sigma_{L,j}\psi_j^{(2)}(0^+), \quad \psi_j^{(2)}(t+\beta/2+0^+) = -\sigma_{R,j}\psi_j^{(1)}(t+\beta/2-0^+),$$
$$\psi_j^{(2)}(2t+\beta-0^+) = -\sigma_{L,j}\psi_j^{(1)}(0^+), \quad \psi_j^{(1)}(t+\beta/2+0^+) = \sigma_{R,j}\psi_j^{(2)}(t+\beta/2-0^+). \tag{10}$$

It is important to notice that $\psi_j^{(1)}(2t+\beta) = -\sigma_{L,j}\sigma_{R,j}\psi_j^{(1)}(0)$, which becomes the standard anti-periodic boundary condition only if $\sigma_{L,j} = \sigma_{R,j}$. Performing the integration over Gaussian variables $J_{i_1 i_2 \cdots i_q}$, we find

$$Z_{\text{SRE}} = \sum_{\{\sigma_{L/R,j}\}} \int_{\text{B.C.}} D\psi_j^{(\alpha)}\ \exp\bigg(-\int ds \sum_{j,\alpha}\frac{1}{2}\psi_j^{(\alpha)}\partial_s\psi_j^{(\alpha)} + \frac{J^2}{2qN^{q-1}}\sum_{\alpha\gamma}\int dsds' f(s)f(s')\Big[\sum_j \psi_j^{(\alpha)}(s)\psi_j^{(\gamma)}(s')\Big]^q\bigg). \tag{11}$$

Next, we introduce the bilocal fields $G(s, s')$ and $\Sigma(s, s')$ following the standard SYK technique [4]. We also redefine the fields from $\psi_j^{(\alpha)}$ to $\psi_j^{[\alpha]}$, as discussed in the main text. The resulting expression reads

$$Z_{\text{SRE}} = \sum_{\{\sigma_{L/R,j}\}} \int_{\text{B.C.}} D\psi_j^{[\alpha]} DG^{[\alpha\gamma]} D\Sigma^{[\alpha\gamma]}\ \exp\Big(-S[\psi_j^\alpha, G^{[\alpha\gamma]}, \Sigma^{[\alpha\gamma]}]\Big),$$

$$S = \int dsds' \frac{1}{2}\psi_j^{[\alpha]}(s)\Big(\delta^{\alpha\gamma}\delta(s-s')\partial_{s'} - \Sigma^{[\alpha\gamma]}(s,s')\Big)\psi_j^{[\gamma]}(s') + \frac{N}{2}\bigg(\Sigma^{[\alpha\gamma]}(s,s')G^{[\alpha\gamma]}(s,s') - \frac{J^2 f(s)f(s')}{q}G^{[\alpha\gamma]}(s,s')^q\bigg). \tag{12}$$

3Here, the summation over indices is kept implicit. The boundary condition now reads

$$\psi_j^{[\alpha]}(2t + \beta - 0^+) = -\sigma_{L,j}\psi_j^{[\alpha]}(0^+), \quad \psi_j^{[\alpha]}(t + \beta/2 + 0^+) = \sigma_{R,j}\psi_j^{[\alpha]}(t + \beta/2 - 0^+). \tag{13}$$

To derive the saddle-point equations, we integrate out $\psi_j^{[\alpha]}$, which yields

$$Z_{\mathrm{SRE}} = \int_{\mathrm{B.C.}} DG^{[\alpha\gamma]} D\Sigma^{[\alpha\gamma]} \Big[ \sum_{\sigma_L \sigma_R} \det(\partial_s - \Sigma^{[\alpha\gamma]})_{\sigma_L \sigma_R} \Big]^{N/2} e^{-\frac{N}{2}\int[\Sigma^{[\alpha\gamma]}G^{[\alpha\gamma]} - \frac{J^2 ff}{q}(G^{[\alpha\gamma]})^q]}. \tag{14}$$

Here, the subscript indicates that the derivative is defined subject to the boundary condition (13). Varying with respect to the bilocal fields, we find

$$\Sigma^{[\alpha\gamma]}(s,s') = J^2 f(s)f(s')G^{[\alpha\gamma]}(s,s')^{q-1}, \qquad G^{[\alpha\gamma]}(s,s') = \frac{\sum_{\sigma_L \sigma_R} g_{\sigma_L \sigma_R}^{[\alpha\gamma]}(s,s')\det[g_{\sigma_L \sigma_R}^{[\alpha\gamma]}]^{-1/2}}{\sum_{\sigma_L \sigma_R} \det[g_{\sigma_L \sigma_R}^{[\alpha\gamma]}]^{-1/2}}, \tag{15}$$

where we introduce $g_{\sigma_L \sigma_R}^{[\alpha\gamma]}(s,s') = (\partial_s - \Sigma^{[\alpha\gamma]})_{\sigma_L \sigma_R}^{-1}(s,s')$ for fixed boundary condition. After solving the saddle-point equations, the partition function of the SRE reads

$$-\frac{\ln Z_{\mathrm{SRE}}}{N} = -\ln\left(\sum_{\sigma_L \sigma_R} \det[g_{\sigma_L \sigma_R}^{[\alpha\gamma]}]^{1/2}\right) + \frac{(q-1)}{2q}\sum_{\alpha\gamma}\int ds\,ds'\,\Sigma^{[\alpha\gamma]}(s,s')G^{[\alpha\gamma]}(s,s'). \tag{16}$$

Finally, as discussed in the main text, fields $\psi_j^{[\alpha]}$ with different $\alpha$ are disconnected under the boundary condition (13). For the SYK model, this implies that the two-point function $G^{[\alpha\gamma]}(s,s') = \delta^{\alpha\gamma}G(s,s')$ is diagonal in $\alpha\gamma$ [5]. Similar relations hold for $\Sigma^{[\alpha\gamma]}$ and $g_{\sigma_L \sigma_R}^{[\alpha\gamma]}(s,s')$. We also have $\det[g_{\sigma_L \sigma_R}^{[\alpha\gamma]}] = \det[g_{\sigma_L \sigma_R}]^4$. Furthermore, in Eq. (15), contributions with $\sigma_{L,j} \neq \sigma_{R,j}$ vanish identically. This follows because

$$\begin{aligned}
\det[g_{-\sigma_R,\sigma_R}^{[\alpha\gamma]}]^{-1/2} &= \int_{-\sigma_R,\sigma_R} D\psi^{[\alpha]} e^{-\frac{1}{2}\int \psi^{[\alpha]}(\delta^{\alpha\gamma}\partial_{s'} - \Sigma^{[\alpha\gamma]})\psi^{[\gamma]}} \\
&\propto \int_{\sigma_R,\sigma_R} D\psi^{[\alpha]} e^{-\frac{1}{2}\int \psi^{[\alpha]}(\delta^{\alpha\gamma}\partial_{s'} - \Sigma^{[\alpha\gamma]})\psi^{[\gamma]}} \prod_\eta \psi^{[\eta]}(\beta + 2t).
\end{aligned} \tag{17}$$

Here, we have employed the fact that $\psi_j^{(1)}\psi_j^{(2)}\mathcal{S}_{j,L}^{(12)}(\sigma_{L,j}) \propto \mathcal{S}_{j,L}^{(12)}(-\sigma_{L,j})$. Since different replicas are disconnected, applying Wicks theorem to Eq. (17) yields zero. Putting all ingredients together, the saddle-point equation (15) reduces to Eq. (14) in the main text.

## NUMERICAL RESULTS WITH SYK$_2$ OR SYK$_6$ PERTURBATIONS

In the main text, we argue that the slow saturation in the low-temperature SYK model with $q = 4$ originates from the soft reparametrization mode governed by the Schwarzian action. Here, we provide further evidence by introducing SYK-like perturbations to the original SYK$_4$ Hamiltonian:

$$\delta H = \sum_{i_1 < i_2 < \cdots < i_{q_V}} i^{\frac{q_V(q_V-1)}{2}} V_{i_1 i_2 \cdots i_{q_V}} \psi_{i_1}\psi_{i_2}\cdots\psi_{i_{q_V}}, \tag{18}$$

where the random couplings satisfies

$$\overline{V_{i_1 i_2 \cdots i_{q_V}}} = 0, \qquad \overline{V_{i_1 i_2 \cdots i_{q_V}}^2} = (q_V - 1)! V^2 / N^{q_V - 1}. \tag{19}$$

We focus on two senarios with $q_V = 2$ or $q_V = 6$. The only modification to the Schwinger-Dyson equation reads

$$\Sigma(s_1, s_2) = f(s_1)f(s_2)\left[J^2 G(s_1,s_2)^{q-1} + V^2 G(s_1,s_2)^{q_V-1}\right]. \tag{20}$$

It is known that, in the low-energy limit, a perturbation with $q_V = 2$ is relevant and drives the system toward the Fermi-liquid fixed point, which is not described by the reparametrization modes [6]. By contrast, the $q_V = 6$

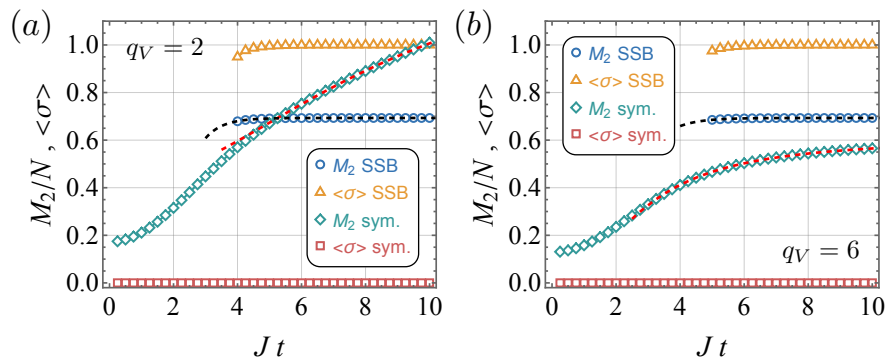

FIG. 1. Numerical results for the SRE in the perturbed SYK model with $q = 4$ and (a) $q_V = 2$ or (b) $q_V = 6$. We fix $V/J = 0.2$ and $\beta J = 4$ in all plots. The dynamical transition occurs at finite $Jt$ for $q_V = 2$, while it remains absent, consistent with the unperturbed case, for $q_V = 6$.

perturbation is irrelevant and can be neglected at low energies, so the Schwarzian action remains valid. Consequently, we expect slow dynamics to persist for $q_V = 6$, but not for $q_V = 2$, even when $V/J$ is small. This expectation is borne out by the numerics shown in FIG. 1, where we set $V/J = 0.2$ and $\beta J = 4$. These results further support the central role of the reparametrization mode in the observed slow dynamics of the SYK$_4$ model.


\fi
  
\end{document}